**This is the "accepted manuscript" version of the following article:**

González-Lluch C., Company P., Contero M., Camba J.D., & Plumed, R. (2016). A Survey on 3D CAD model quality assurance and testing tools. *Computer-Aided Design.*

which is available at: http://dx.doi.org/10.1016/j.cad.2016.10.003

**Title: A Survey on 3D CAD Model Quality Assurance and Testing Tools**

## 1. Introduction

According to the SASIG Guidelines for the Global Automotive Industry, "good product data quality means providing the right data to the right people at the right time [1]." Because the design field is so inherently rich in information, it is necessary to implement systems that can efficiently and securely track, control, manage, and share these data.

Product data is an umbrella term that includes many different types of information. In this paper, we specifically focus on 3D CAD models. In this context, CAD model quality is a key concept, as the quality of a manufactured product depends on the quality of its design processes, which then depend on the quality of their data [2]. High quality product data is essential, as low quality often delays product development, and can negatively impact the overall quality of the final product [1,3].

In this paper, we provide an examination of CAD Model Quality Testing (MQT) tools and techniques, which are used in situations where CAD models contain errors or anomalies (in most cases, unnoticeable to the user) that can hinder simplification, interoperability, and/or reusability. Although in academic circles, Modeling Quality Testing is considered solved by a number of scholars, it has been reported that the implementation and practical application of these solutions in industrial settings is far from trivial [4]. Therefore, Model Quality Testing remains an open problem.

A number of MQT tools such as CAD Checker by Excitech [5,6] are designed to validate the correctness of 2D CAD drawings. While certainly important in industries that have not yet moved to a paperless workflow paradigm based on 3D virtual models, these tools are out of the scope of our study. Instead, we focus on scenarios where 3D CAD models are used as the main vehicle for the delivery of product information and as the central pieces around which downstream CAD/CAM/CAE applications are structured. For example, design engineers involved in Finite Element Analysis, Rapid prototyping, Numerical Control, and Data Exchange activities often need complete, correct, and clean CAD models, so derived models can be efficiently generated while ensuring product quality and minimizing production costs and time to market.

This paper defines a new taxonomy of issues related to CAD model quality. For validation purposes, the taxonomy was used to classify currently available MQT tools, thus determining which aspects of quality are reasonably addressed, and which remain open problems.

The paper is organized as follows: section 2 describes the basic terminology used in CAD Model Quality Testing in relation to the underlying structure of the domain; section 3 presents a systematic taxonomy of the techniques, which is connected to the commercially available CAD Model Quality Testing Tools discussed in section 4. Finally, section 5 concludes the paper by highlighting open research domains.

The paper is extensive as it includes a survey of quality problems in CAD models, introduces a new taxonomy for classifying these problems, and also includes a survey of QMT products.

Although this may seem far too much at once, we strongly believe that a global analysis was required before subsequent sectorial analysis can be efficient. This is because the paper argues against the general believe that MQT tools is an academically solved problem and only incremental improvements are required. Instead, the thesis of this paper is that current MQT tools are mostly aimed at homogenizing the vast amount of documents produced and shared by large OEM's, and thus are primarily aimed at preventing easily solvable low-semantic level mistakes and incoherencies, while we foresee that they will only become valuable for other market segments (like SME's) if document homogenization ceases to be prevalent over conveying design intent.

## 2. Terminology

*Product data* refers to all data involved in the design and manufacturing of a product [7]. *Product Data Quality* (PDQ) is a measurement of the accuracy and appropriateness of these data combined with the timeliness with which they are provided to the stakeholders who may need them [1,8].

A simple product data tool is the *native modeler*, which is the CAD system used to create a *native model* [9], also called *master model* by the Automotive Industry Action Group (AIAG) [2].

A *CAD model* is a mathematical *representation* of a product used to explain and predict the behavior of such product before it is built. *Representations* are organized collections of associated data elements collected together for specific uses [10]. They are the data sources for procedures that compute useful properties of objects [11]. A CAD model is stored in a *document*, which is a fixed and structured amount of information that can be managed and interchanged as a unit between users and systems [12].

CAD models are represented according to languages, which must conform to a certain representation scheme. Two representation schemes currently dominate most Mechanical CAD systems: explicit and procedural. Representations are *explicit* (also known as declarative or evaluated) when their details are immediately available without the need for any calculations. Representations are *procedural* (also known as generative or history-based), if they are described in terms of a sequence of procedures (which may include the solution to constraint sets). Finally, *hybrid* representations result from combining both explicit and procedural representation methods [13].

However, representation of geometrical entities is subject to another source of potential discrepancies. Most geometrical entities may be defined by different sets of parameters. This fact is acknowledged in section 4.2.6 of ISO 10303-42:1994 [14] where in the case that an item of geometry could be defined in more the one way, a "preferred form" o "master representation" should be nominated.  It is recommended that parametrization, domain and results of evaluation to be derived from the master representation. In a note at the end of this section it gives a clear warning: "Any use of the other representations is a compromise for practical considerations".

Modern product data standards such as ISO 10303 follow the ANSI/SPARC three-layer architecture for database systems. This means a clear distinction between the data models (applications and logical layer) and the file format for data exchange (physical layer defined as in part 21) [15]. This is important when considering *downstream applications* (such as CAM and CAE systems) that do not support the native modeler [9,16] as their input and require a direct translation or a neutral format. In this context, master models are also called *primary views*, and exported o derived models, *secondary views* [17-19].

*Strategic CAD knowledge* is the ability to recognize that design choices made during the development of the master model may drastically determine how easy or how difficult it is to perform subsequent design changes to the model [20-22]. Naturally, incorrect or inefficient master models should always be avoided, but even when effective strategies are used, master models may still contain errors. It is in this type of situations where *Model Quality Testing* is required.

It is usually the secondary models which need cleaning up, as a result of data transfer problems. But discrepancies between geometry and functionality result in errors in the higher levels of quality of CAD models. It follows if we accept that semantics focuses on the relationship between signifiers (elements of geometry in our case), and what they stand for (the functionality of one particular shape, used to solve a certain design problem). Therefore, master models may also need cleaning up to remove their semantic level errors, if strategic CAD knowledge is pursued. Thus, we agree that CAD *Model Quality Testing* is an activity that involves identifying "dirty clean-up problems" in a master CAD model. *Model Quality Technology* enables designers to identify, locate, and even resolve model integrity problems before the file leaves the CAD system [2]. Early advances in *Model Quality Assurance Systems* were summarized by [7].

Resolving model integrity problems requires *model repair/healing*, which entails making slight adjustments to the geometry to remove anomalies [7,8,23]. *Healing* and *repairing* involves processing a model with undesirable artifacts and creating a new version that is free from these errors [24]. There is no general agreement on whether model healing is beneficial as (1) repairing secondary views without propagating the changes to the primary view compromises data integrity, and (2) repairing a local problem in one area of the model may create new local problems elsewhere or affect the design intent of the whole part [7,23]. This controversy is out of the scope of our paper, but certainly a topic that needs to be considered in related studies. It is generally agreed that model repair/healing is useful for homogenizing vast amounts of documentation used by OEM's and their network of suppliers. However, if focus is on detecting higher level errors to further improve quality of CAD models, homogenization becomes just a small part of the problem. In this context, where homogenization is not prevalent over design intent, some repair strategies may be hiding or even inverting the original design intent.

In order to describe and classify the types of quality loss that can be automatically detected and repaired, it is important to further clarify two aspects: the *type of representation* that describes the model, and the *type of change* that causes the loss of quality.

## 2.1 Types of representations

Two representation schemes currently dominate most Mechanical CAD systems (MCAD): Boundary Representation (B-Rep), and history-based parametric feature-based models [19,25].

This distinction is not always accepted as, actually, a feature-based CAD model is made of two interconnected representations: B-Rep and feature-based. But we follow the item 3.7.28 of the standard ISO10303-108:2005(E) [13], which defines procedural models "as opposed to an explicit or evaluated model which captures the end result of applying those procedures". An explanatory note follows: "NOTE. Although procedural models are outside the scope of this part of ISO 10303, they are defined here to make an important distinction between two fundamentally different modelling approaches. The present resource is intended to be compatible with ISO 10303-55, which provides representations for the exchange of procedurally defined models." Furthermore, when explicit models are defined (item 3.7.16) another explanatory note is added: "In the case of product shape models, an explicit (or evaluated) shape model is a fully detailed model of the boundary representation or some related type, as defined in ISO 10303-42. More specifically, an explicit model is a model that is not of the procedural or hybrid type, which may contain little or no explicit geometry". Finally, the item 3.7.19 of the standard ISO10303-108:2005(E) defines hybrid models, which are useful because one of their representations is better for some actions (like speeding model render), while the other is better for other purposes (like answering geometrical queries posed by the user).

B-Rep representation is a particular type of *explicit representation* where suitable sets of connected geometric elements are used to represent the vertices, edges, and faces that define the boundary of a solid [26]. *Geometric information* defines the exact shape and spatial position of the elements, whereas *topological information* defines the links between the elements [27]. Some authors have further classified B-Rep representations as B-Rep, faceted B-Rep, and boundary curve-based [28].

Generally speaking, B-Rep models consist of patches. B-Rep *patches* are portions of curved surfaces bounded by *curved contours*, which are pieces of curves delimited by *vertices* (which lie at points). There seems to be a terminological confusion in the literature, as patches are sometimes called faces. Some authors consider a face as any bounded portion of a surface (it may be planar, cylindrical or adopt any curved form) while other researchers claim that faces must always be planar (they are only faces in the particular case where a flat contour delimits a region of a plane). For example, a can is modeled by three patches: one cylindrical patch and two faces delimited by circles. Then, the *shell* is a set of connected patches/faces. But the literature also differentiates shells as piecewise-algebraic surfaces (made of algebraic pieces) or piecewise-linear surfaces (made of linear pieces).

On the other hand, there is a problem regarding the sizes of patches/faces. B-Rep patches are common but inefficient to render to a computer screen and difficult to process in physical simulations [29]. First, each shape requires its own particular algorithm to simulate its behavior, and second, this information is unavailable in models generated from point clouds [30] or sets of scanned images [31]. In other words, there is a need to convert the trimmed surfaces of B-reps into a form that removes the gaps between adjacent faces and allows a homogeneous representation of shape [32]. In these cases, the best solution involves discretizing the model into a *mesh* of smaller elementary mesh cells of mesh elements (usually triangles) that approximates a geometric domain [33]. According to [34], a mesh is an arrangement of cells with connectivity between the cells defined by the possession of common cell faces or cell edges. A cell is a manifold of dimensionality one or higher that is a part (or the entirety) of a mesh. In this paper, we use the term *face* to refer to a mesh cell. Thus, the cylindrical patch of the can from the earlier example may be subdivided into a mesh of an adjacent slender quadrilateral faces (that approximate the cylindrical surface as an n-sided prism). Computer generated data produces *tessellated* meshes, while scanned data produces *digitized* meshes [35].

A recent survey by Shimada discusses current trends and issues in meshing and geometric processing for computational engineering analyses [31]. Researchers Attene et al. analyzed common mesh defects, and surveyed existing algorithms to process, repair, and improve meshes [35]. In our paper, we distinguish between B-Reps (made of patches), and *meshes* (made of smaller homogeneous faces). *Subdivision surfacing* is a method where a coarse piecewise-linear polygon mesh is recursively refined by subdividing each polygonal face into smaller faces that better approximate the desired smooth surface. Alternatively, mesh reduction techniques are applied to dense triangle meshes to build simplified approximations while retaining important topological and geometric characteristics of the model. The reduced mesh is interpolated with piecewise-algebraic surface patches which approximate the original points [30].

Solid models are defined by the closed shell that encloses its material. In this context, the term *closed* means that the surface of the model divides the entire space into disjoint internal and external volumes [36]. Closed B-Rep and meshed models provide a complete representation of a solid shape, but do not save the details on how the shape was created [37].

Alternatively, *procedural modeling* techniques such as history-based parametric feature-based modeling create 3D models from sets of rules [38]. Procedural models have the advantages of capturing all or part of the design intent and being easy to edit [39]. Procedural approaches to create 3D CAD models include feature-based design (FBD) and constraint-based modeling [40,41]. The recent emergence of feature-based downstream applications such as Numerical Control (NC) machining [42] is favoring the already dominant position of feature-based master models.

We observe that the representations described in this section may result from interaction with the user or from automatic algorithms that shift from a low semantic representation up to a higher one. Reverse engineering (which is aimed at extracting knowledge or design information from actual products in order to produce their CAD models) is a complex problem by itself and, as such, is out of the scope of this paper. Only the final output is classified in this study.

## 2.2 Types of changes

There is general consensus on the types of changes that can be performed to a master model: simplification, interoperability, and reuse.

*Simplification* means converting accurate, highly complex models into simpler models that retain the important details and eliminate the irrelevant ones [16]. Reducing the geometric complexity of the master model at various *levels of detail* (LODs) is desirable and useful so valid simulations can run at a reasonable cost. Feature-based *multi-resolution solid modeling* strategies are helpful for creating secondary views (i.e. different versions for different needs [43,44]). In some cases, users may want to remove the "know-how" from simplified models used as final deliverables for fear of losing intellectual property (as argued, for instance by [45]).However, it would be inefficient to remove the "know-how" entirely from all simplified models created for internal use.

*Interoperability* refers to the level up to which a master model can be accurately transferred from one modeler to another [2,9]. Interoperability is concerned with the resolution of semantic differences between similar constructs in different representations and the inherently different tolerances within the geometric algorithms used at the core of different CAD modelers [46,47]. Interoperability encompasses the exchange of CAD models between different CAD systems and to downstream applications [48]. Early attempts to solve product information exchange problems were described by [49,50]. Interoperability is particularly important in collaborative design environments, where CAD/CAM/CAE [51] and information and knowledge sharing tools are commonplace [52].

Model *reuse* involves modifying a master model in the native modeler so it can be utilized in other situations. Reuse can be performed at different levels ranging from selecting library items to adapting existing designs [53]. In general, two types of reusability can be defined: (1) editing a model to produce a redesigned version of the current object (*instantiating)*, and (2) cannibalizing old designs to use "spare parts" (*version modeling)* [54,55]. Modifications are frequent, even for instantiated models, so the quality of reusable models must be particularly high, as they need to reliably allow for modifications while maintaining the original design intent. A related problem, model retrieval for reusability, has also received some attention by the research community [56-58].

To note that the coupling of the three different types of changes may influence on the quality of the CAD model [45]. However, the current classification is limited to consider the impact of each action separately, which is still a highly challenging goal.

## 3. Taxonomy of CAD model quality

In order to understand CAD model quality issues, Pratt [23] emphasized the importance of the *semantic study*. By further developing this idea, authors Contero et al. [18] defined three *levels of quality* to classify CAD models: *morphologic*, which relates to the geometric and topological correctness of the CAD model; *syntactic*, which assesses the proper use of modeling conventions, and *semantic/pragmatic*, which focuses on the CAD model's ability for modification and reuse. Other authors classified quality issues by the *type of representation* [37,42].

The taxonomy presented in this paper defines a "frame" with three levels of CAD quality for each type of representation, and then identifies the artifacts that may hinder simplification, interoperability, or reuse of the model. The general levels of our taxonomy are shown in Fig.1. In the next sections, we argue that each semantic level can be mapped (to a certain extent) to a particular type of change: morphologic to simplification, syntactic to interoperability, and semantic to reuse.

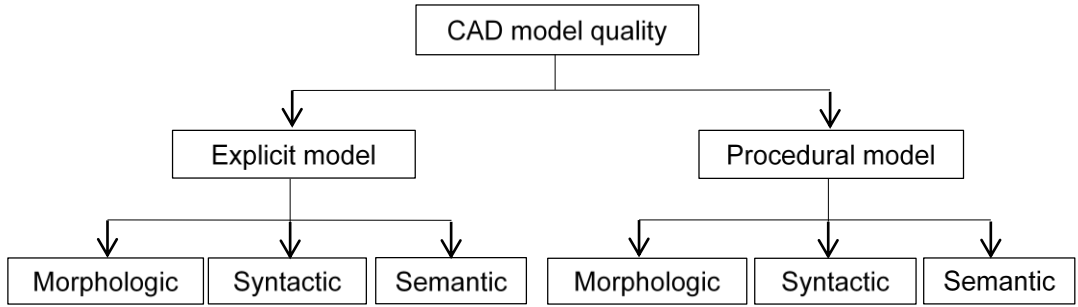

Fig. 1. General levels of CAD Model Quality

Figure 1 highlights that classifying only by representation type or only by the semantic level of the failures are both incomplete views. However, it omits the big differences inside each category—for instance, B-rep and mesh models are very different explicit models—as adding more details (like sub-categories of explicit or procedural models) would mask the main proposal of a "frame" that merges the two relevant dimensions.

In the following sections, each of the six types defined in Fig.1 is further subdivided in *sub-types* (represented by horizontal boxes with horizontal text). For each subtype, some of the most representative *errors* found in the literature are included to better illustrate the scope of the sub-type (vertical boxes with vertical text).

We note that some readers may be reluctant to perceive some of the following errors, since it is quite commonly accepted that some models have a very concise, simple, common data structure, where it seems that no room for errors is left. However, all languages may be used to convey the wrong message, or to incorrectly convey the right message. For instance, syntactic and semantic errors in mesh models may be hardly perceived. But mesh models usually prioritize regularity of the mesh (as they are usually aimed at visualization or Finite Elements Analysis), and thus may introduce quality errors in some critical CAD/CAM properties, like small variations in the exact position of the holes of a flange.

### 3.1 Morphologic errors in explicit CAD models

The morphological level of quality for B-Rep and mesh representations has been widely studied, as morphologic errors may affect simplification, which is critical for downstream applications. For example, authors Bøhn and Wozny considered *punctures* (which produce open shells), *inconsistent facet-orientations* (which create uncertainty in terms of what side of the shell contains material), and *internal features* (non-manifold elements within the shell that create ambiguities when solidifying) [59].

Gu et al. distinguished geometric errors (those that produce solids that are topologically valid, but potentially difficult to simplify by downstream applications) from topological errors (those that result in invalid topologies) [9]. By observing the effect of errors on a model, the authors concluded that the former type (topological) includes *realism* errors, including: (1) singularities characterized by sudden changes in normals or tangents, (2) abnormally small or uncommon volumes or holes, and (3) abnormally small distances between edges or vertices. They subdivided topological errors as: (1) *Accuracy* errors that consist of excessive geometric gaps between topologically connected elements (like one vertex that falls outside a tolerance volume which surrounds the intersection between the edges that should share it), and (2) *Structure* errors that occur when any topological element of a model, including vertices, edges, faces, and shells, is undefined or incorrectly linked [9].

Researchers Yang et al. [28] identified six types of CAD model errors in explicit representations: tiny faces, narrow regions, non-tangent faces, narrow steps, sharp face angles, and narrow spaces. The catalog of error types was updated by Chong et al. [60], who used the source of the errors to identify six types of geometric errors (non-manifold mesh, small gaps and overlaps, large gaps and overlaps, duplicated or undesired elements, surface orientation and sliver surfaces), and two types of topological errors (holes, and missing parts/surfaces).

More recently, Attene et al. [35], prioritized geometric realization to distinguish between geometric issues (which, in addition to punctures, include intersections, near degeneracies, noise, and feature chamfering/aliasing), local topology flaws (non-manifold mesh), and global topology flaws (inconsistent orientation and wrong genus).

Several error classifications have been proposed. The main difference is whether errors should be considered geometric or topological, which depends on whether the classification considers what caused the error (source), how the error affects the whole object (scope), or how the error manifests (visibility). Gaps and holes break the connectivity (i.e., they change the topology), but they usually come from imperfect geometric calculations (source), and are usually detected by counting connections (visibility). A similar problem applies to the distinction between local and global errors. For instance, a local displacement of a vertex may cause a global loss of symmetry.

Our classification targets the source of the artifacts and their scopes, while paying less attention to their visibility (which is oriented at defining procedures and algorithms that detect and repair the errors). Assuming this approach, errors that violate the design intent but affect to isolated elements are local. They are topological, if the isolated elements are the incorrect ones; and geometrical, if the isolated elements interact with other elements (or with references) in the wrong manner. For instance, parallelism is a geometrical property, thus, two faces that fail to be parallel determine a geometrical problem. If the lack of parallelism of those faces causes another error to appear elsewhere (i.e. propagates), then the original error is global. Otherwise, it is local, despite the size of the faces. Therefore, we distinguish:

- Local geometric issues are generally due to isolated accuracy errors, including punctures and overlaps.
- Global geometric issues are noise caused by repetitive accuracy errors.
- Local topology flaws include non-manifold meshes, unrealistic shapes, and abnormally small or uncommon elements.
- Global flaws in topology include inconsistent orientation and wrong genus.

The length to width aspect ratio plays an important role when examining punctures. Length and width are similar in large gaps or *holes* (those having three or more borders). Small gaps are slender and include only two borders [61]. We consider *T-joints* as a particular type of gap where the number of edges and vertices on opposite borders are different [60,62,63]. Transversal penetrations (or clashes) and intersections are the complementary of holes and gaps. They can be classified according to the type of contact curve: closed (for interpenetrations) and open (for intersections). Finally, both small gaps/overlaps and rendering inaccuracies may cause visual artifacts (known as *cracks)* between adjacent faces, which are not considered in this paper, as they affect the visualization quality of the model [64,65].

The most common global geometric issue is the repetitive noise that results from data corruption when processing digitized meshes. This noise distorts most of the elements of the mesh, producing undulations, such as ripples and creases. The problem is common in digitized meshes, but not as common in B-Rep models.

In order to classify local topological errors, it is important to review some concepts. A mesh is *two-manifold* if it is homeomorphic to a disk in the neighborhood of every point (i.e., resembles a two dimensional Euclidean space near each point). We can further sub-divide non-manifold models as those produced by *under-connected* elements (also called isolated, dangling, or naked elements), or *over-connected* elements (singular elements). For instance, when more than two polygons share a common edge, then such edge is said to be singular, complex, or non-manifold.

According to Gu et al. [9], unrealistic shapes are linked to sudden changes in normals or tangents, and can be further subdivided into appendage-volumes (tiny features that protrude), knife-edges (two faces that join at a sharp edge), and sliver-cracks (sharp wedge-shaped cracks between features). Abnormally small or uncommon volumes or holes include sliver-faces, micro-slots, etc. A sliver face is a face with a high aspect ratio, whereas a spike is a region with a high aspect ratio inside a face [66]. Abnormally small distances between edges or vertices can also be the cause of errors.

Global flaws in topology include inconsistent orientation and wrong genus. The latter is also called topological noise [67] and is due to the finite precision of digitization tools that alters the number of handles and tunnels in the geometry [68].

Anisotropy and directionality are particularly important properties in engineering simulations [31]. They are indirectly imposed by Delaunay meshes (which maximize the minimum angle of all the angles of the triangles in the triangulation), or sliver-free meshes (a *sliver* is a nearly flat tetrahedron in a tetrahedral 3D mesh [69]). Consequently, their absence is detected as a potential error by some MQT tools.

Morphologic errors in explicit models are summarized in Fig. 2. Horizontal boxes represent the realm of the errors. Vertical boxes identify common errors linked to those realms.

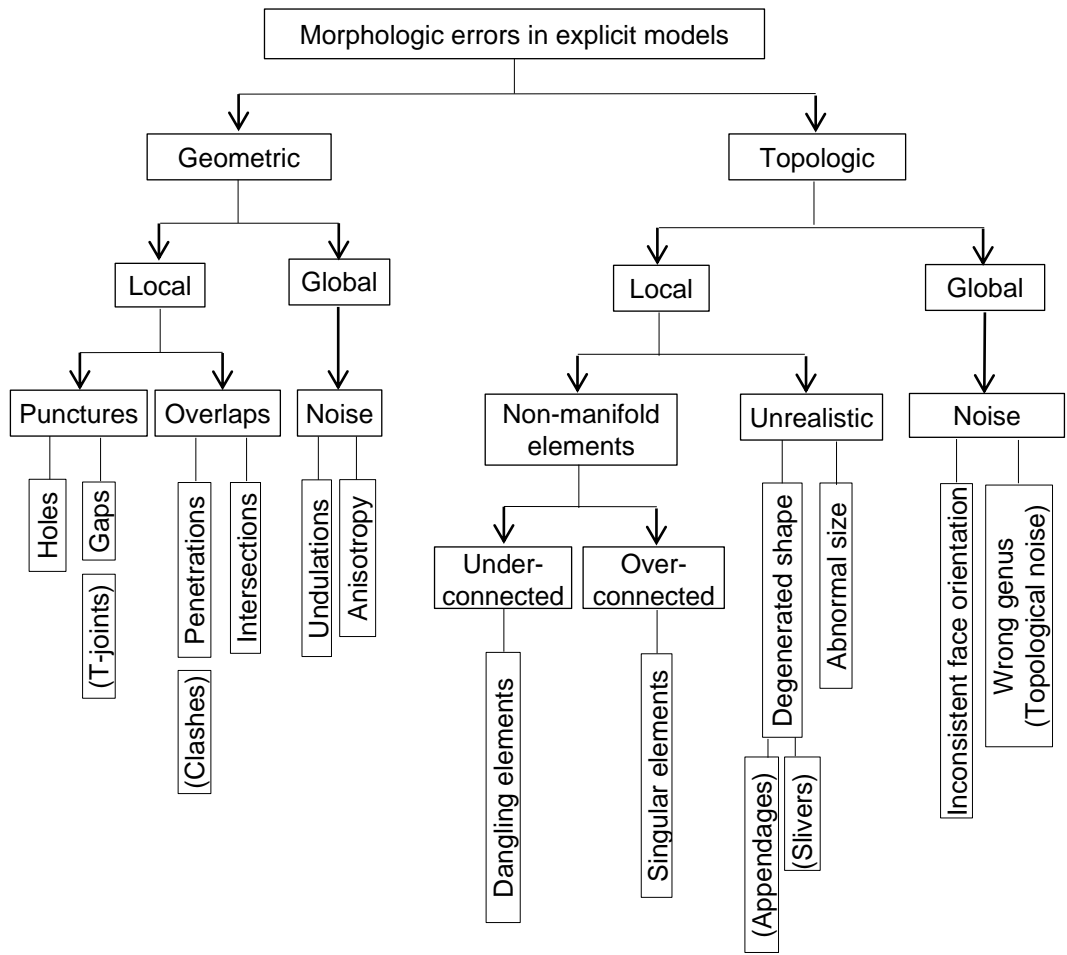

Fig. 2 Morphologic errors in explicit models

A general technique to prevent geometric errors is robust computation, which introduces knowledge at every possible level to guide the computation [70]. Various strategies can be used to automatically repair the remaining errors. For example, approaches exist to heal meshes with holes [33,71]. Gaps can be stitched by physically moving vertices (i.e., changing their coordinates), although this change introduces an error in the order of the resolution tolerance [72]. *Geometric hashing* finds particular instances of objects in a given scene, even if they are partially occluded [73]. Hashing techniques heal gaps by assuming that one border is a slightly deformed sub-curve of the other. Then, they find the translation and rotation of the sub-curve that yield the best least-squares fit to the appropriate portion of the larger curve, and construct new patches that fill the remaining holes (typical patches are triangles obtained by the 3D minimum-area triangulation technique) [74]. Alternative approaches, classified as "loyal" to the input model, have also been described in the literature [75].

Local flaws in topology can be solved by most commercial applications. These flaws are detected through indirect cues like counting connections, or using angles and sizes to diagnose various artifacts [9].

Wrong genus can be solved by Topology Denoise Technologies [76]. However, manual healing is often required to ensure anisotropy and directionality, which are not effectively handled by commercial mesh generators [31].

## 3.2 Syntactic errors in explicit CAD models

Syntactic errors are often the source of interoperability problems that appear because the native model and the downstream applications use different data structures.

Although there are different strategies to address interoperability [77], the most practical solution requires *mapping*, which involves univocally associating every geometric description from the original representation to a fully equivalent geometric description in the image representation [78]. Mapping between data structures may be *direct*, or use a *neutral* exchange format [9]. Direct Translation is the ability to read one CAD format and write the information to a different format. The fundamentals of the direct mapping style are available in [79]. Most of the solutions in this field are practical, but commercially protected (although some became publicly available [80]). The alternative is saving CAD data to a neutral format, so a second CAD system can read it.

Direct translation is advantageous when systems share the same *kernel* (the *geometric engine* that stores and organizes the basic geometric shapes and model topologies), as these CAD systems can directly share modeling data recorded in the kernel's native format [46]. Nevertheless, kernel-level data exchange is limited by the different "flavors" developed by manufacturers, which are often protected for commercial reasons [81].

Translation problems due to neutral files can be subdivided in *inaccuracy*, *inconsistency*, and *loss of semantics* [46]. The former two problems are explained in the next paragraphs. Loss of semantics is described in the next section.

A model that is inaccurately translated becomes *disintegrated*, as its entities become disconnected or fall outside the required tolerance (as a result of moving them from one system to another that uses tolerances that are too small or too large) [46,82-84]. The underlying problem is the inability to maintain complete internal consistency between topological and geometric information. This problem is known as the *geometric accuracy problem* [85,86] and is caused by computational errors in geometric calculations [23]. It has been stated that precision is particularly necessary for NURBS surfaces in large models with fine details [87]. Yang et al. [88] identified disintegration errors in the IGES and STEP neutral formats.

For the sake of brevity this paper does not consider the inconsistencies caused by the lack of standardization. However, (a) failure to meet standards may result in some data not being visible or being incorrectly interpreted [3], and (b) there are mapping incoherencies due to the shift from old technical drawing standards [89] to 3D model-based standards such as ASME Y14.41 [90].The only standardization topic we consider is that *standardizing neutral formats* helps prevent inconsistent mappings of 3D CAD models. In this regard, the most relevant standard neutral formats are IGES and STEP.

"Initial Graphics Exchange Specification" (IGES) was first published in 1980 as "Digital Representation for Communication of Product Definition Data" by the U.S. National Bureau of Standards (NBSIR 80-1978). STEP (Standard for the Exchange of Product model data, ISO 10303) [91,92] includes two standard formats that are widely implemented: AP 203, maintained by PDES, Inc. and AP 214, maintained by ProSTEP iViP and SASIG2.

Some inconsistencies have survived standardization. First, some standards are sectoral to certain industries. For example, AP 203 can be used to transfer purely geometric models with a high degree of success [37]. It is primarily supported by the aerospace and defense industry, but AP 214 is the preferred format in the automotive industry [93]. Since standards are a subset of all the information that is relevant to a diverse number of CAD systems, they can only represent the geometric information that is common to all systems, [94].

Second, standardization may be unintentionally open to interpretation. For instance, IGES translators never became fully standardized, and thus their rules vary greatly from vendor to vendor, which produces inconsistencies [46].

With regard to mapping scope, it is incomplete as it fails to support *legacy* and *domain*. The strategy to support legacy is *durability*, which requires compatibility (consistency along time) between CAD data of different versions of the same system [84]. *Domain mismatches* are failures to support the full domains from different understandings and assumptions. The most common domain mismatch is *incomplete domain*. The most acclaimed STEP AP214 translators have only implemented conformance classes 1 and/or 2, which are essentially identical to AP203. Most commercially available AP214 translators address only the AP203 "look alike" conformance classes (i.e., AP214 cc's 1 & 2) [46].

Syntactic errors in explicit models are summarized in Fig. 3. Horizontal boxes represent the realm of the errors. Vertical boxes identify common errors linked to those realms.

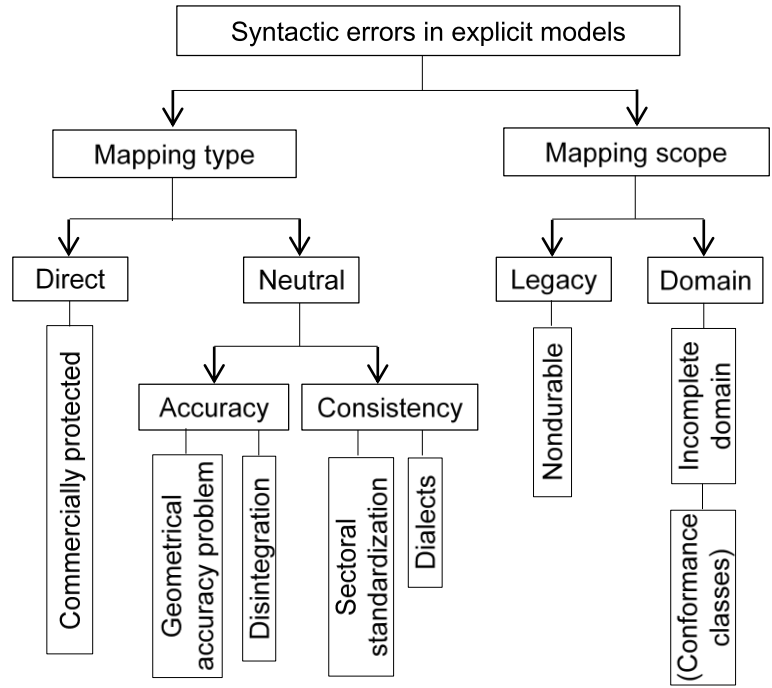

Fig. 3 Syntactic errors in explicit models

Full implementations of neutral formats are still required to prevent miscommunication from incomplete domain mappings. Adoption of neutral formats by Small and Medium Enterprises is also pending, although neutral formats for explicit CAD models lack the semantic level that would motivate those enterprises to adopt them, as explained in the next section. The problem is particularly aggravated for SME's that are not connected to OEM's, because they lack the motivation to adopt those strategies, and the stimulus of long production series to offset the added costs.

### 3.3 Semantic errors in explicit CAD models

As long as semantics focuses on the relationship between signifiers (geometry elements in our case), and what they stand for (the functionality of one particular shape, used to solve a certain design problem), discrepancies between geometry and functionality result in semantic level errors in CAD models. Clearly, there is no single source for those errors. When and how they come to be introduced in the model depends on both the type of modeling and the flow of the design data.

Reusability implies transferring design intent and manipulating the geometry of an existing CAD model, so it can be used in new situations. Reusability refers to the ability to make changes as if the model was created in the receiving feature-based CAD system. Efficient and adaptive reuse is only feasible if design intent and design rationale are available to a certain extent, as it occurs in procedural models [95]. While it might be easy to detect problems in B-Reps or meshes, it is often impossible to automatically solve them if the designer's intent is lost [51,59].

When reuse of explicit models is the only option, reverse engineering strategies (also called *refeaturing*) must be used to find features [96]. In this process, "plain" B-Rep models are "enriched" by shifting to "hybrid" models that include part of the design intent. Different strategies have been used to enrich explicit models. Beautification is perhaps the most successful. A beautified geometric model is a modification of the input model that incorporates appropriate symmetries and regularities [47,97,98]. For instance, the Finite Element (FE) mesh of a symmetric part will not inadvertently be made non-symmetric during editing, if a particular approach for detecting design intent has enriched the model [99]. Declarative feature recognizers are the most recent advance for refeaturing [100].

Refeaturing may include restoring the face structure of a B-Rep from a triangular mesh [101,102], recovering features [103,104], and finding geometric constraints [105]. Currently, expert user guidance is required (setting tolerances to define "likelihoods"), not all types of features can be detected (free-form surfaces and general sweeps are excluded), and some are prone to errors (due to the noise and incompleteness of measured data, and the numerical nature of the subsequent algorithmic phases) [105].

The sampling process (also called acquisition or re-meshing) may produce a particular type of noise that affects tessellated or digitized meshes. It produces irregularly triangulated chamfers (*aliasing artifacts* that appear when restrictions in the location of vertices prevent them from coinciding with sharp edges and corners in the model), thus forcing sharp edges and corners to be restored by feature sharpening approaches [35].

The lack of suitable methods to detect reference systems, as well as suitable datums to skeletonize the model, may prevent finding constraints (both directly or as a result of misaligned reference systems and undetected datums). Some steps have been taken in this direction [99,106].

There are geometric simplifications that need to be performed in preparation for Computer-Aided Engineering (CAE) processes: geometric feature removal and *dimensional reduction* [31]. The former is also known as *defeaturing*, as some features that convey functionality (fillets, chamfers, holes, fasteners, etc.) are removed from the geometry to analyze certain behaviors of the part [107,108]. A common example of the latter type is converting a thin-walled solid into a shell by extracting its mid-surface and meshing it [109].

Dimensional reduction usually requires expert manual adjustments. Additionally, boundary conditions which are critical for assembly analysis are frequently lost [31].

Defeaturing a B-Rep model produces a "wound", or hole that needs to be covered by a replacement surface [110]. Defeatured models may produce different analysis results from those obtained with the original fully-featured model, which may result in defeaturing-induced analysis error and sensitivity modification [111-114]. In front of the argument that there are no defeaturing semantic errors (since results of a deliberate defeaturing can hardly be called errors, while if the process is done incorrectly, that's just a software bug, not a semantic error), we argue that from the point of view of a computer scientist (a programmer), producing the wrong model is a software bug. But, from the point of view of a product engineer (a user), the imperfect software (the only available) outputs a semantic error in the defeatured model.

Semantic errors in explicit models are summarized in Fig. 4. Horizontal boxes describe the realm of the errors. Vertical boxes identify common errors linked to those realms.

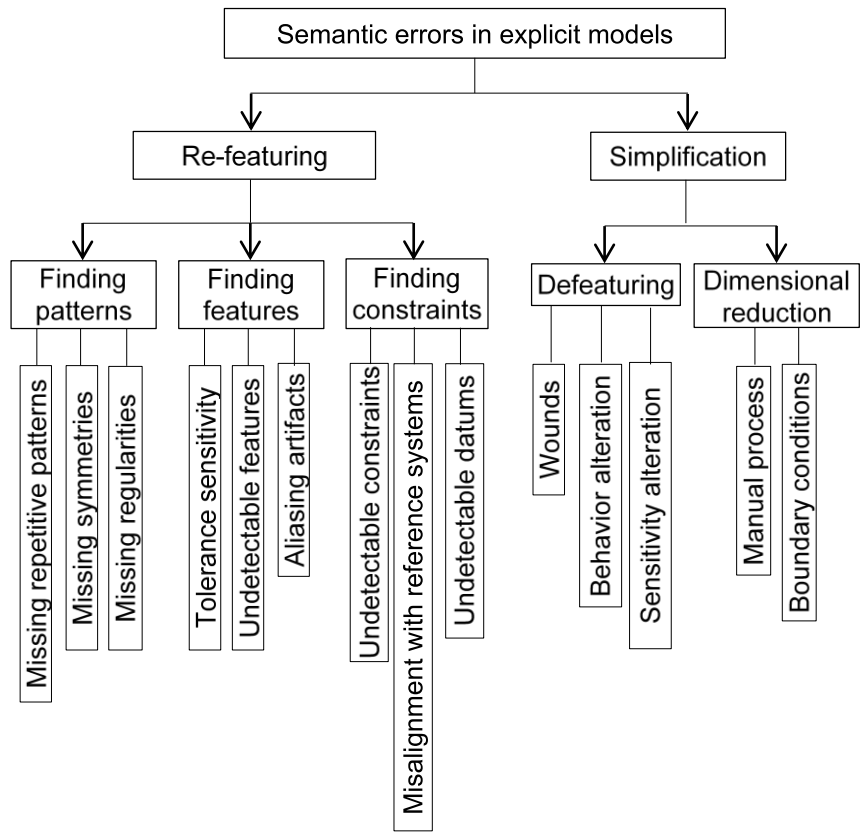

Fig. 4 Semantic errors in explicit models

### 3.4 Morphologic errors in procedural CAD models

The morphological level is not critical in procedural models, as it is automatically tested by most CAD applications (the term "not critical" means important while also solvable to a great extent). These systems permanently check the correctness of the model trees (including constrained profiles, sweeping operations, features and datums), and warn the user about any incoherencies (although careless users may certainly ignore these messages).

In this context, Company et al. developed a classification that includes six dimensions of quality in CAD models: validity, completeness, consistency, conciseness, simplicity, and conveying design intent [115,116]. Validity relates to the morphological level. A model is *valid* when it can be retrieved (it is not missing) and used (it is free of errors).

Sources of morphologic errors include kernel malfunction [70], and inappropriate modeling practices, which may result in subtle variations in shape and size, or inconsistencies that do not cause errors or warning signals in the original model, but may hinder reusability. Legal, and apparently simple, modifications may certainly produce unexpected crashes.

Researchers Yang et al. [28] advocate for a hybrid method to solve these failures. They initially identified six types of CAD model errors in explicit representations (by using the approaches described in section 3.1): tiny faces, narrow regions, non-tangent faces, narrow steps, sharp face angles, and narrow spaces, and used procedural information in the model tree to trace these errors back to the modeling operations. The operations are then reformulated until the errors disappear. The interdependency of feature commands is analyzed to find the best reformulation. However, the shape may occasionally become distorted or even collapse if only the sequence of feature commands is modified. Therefore, the authors also analyzed parametric data to find mutual dependencies between parameters and constraints of different feature commands.

Huang et al. detected problems that affect the NC machining of a part [42]. These failures are similar to the unrealistic shapes described in Fig. 2 and fall in the range of the *completeness* dimension by Company et al. [115], which measures how well the model replicates the actual shape and size of the part. Errors include missing elements as well as those elements that are incorrectly created as a result of kernel malfunction or inappropriate modeling strategies.

Errors caused by ambiguous definitions of procedural models have also been studied. References to entities defined during the design process may be incorrectly reevaluated. Ambiguously defined datums—mainly those defined implicitly ("on the fly")—may inadvertently "switch" after editing the model (e.g. a blend applied to one of the two edges that result from cutting a round slot across a rectangular block [117]). This problem is known as *topological naming* (when names use only topological information) or *persistent naming* (when other types of information are used) [118-122]. Furthermore, Marcheix distinguished between persistent naming of atomic entities (such as vertices, edges or faces) and aggregates (such as sets of faces) [123].The evolution of the approaches used to solve this problem was summarized by [28] and [123], while authors [122] included a detailed taxonomy of persistent naming.

Seemingly well-constrained profiles may sometimes be compatible with different geometric or topological solutions, which are not necessarily compatible with design intent. They appear because most of the problems of the modeling with constraints technique described by Anderl and Mendge [124] are still unsolved. For instance, an arc at a corner of a rounded polygon that produces a blend as a result of sweeping a profile should be tangent to the two lines converging at the corner. However, after a number of changes, internally tangent arcs may be incorrectly selected by the kernel as equally valid solutions [118]. This type of failure may also occur between a modeling operation and its parent modeling operation, when the child operation is relocated to a new location in the design tree (from the standpoint of the Boolean rules that govern Constructive Solid Geometry) and the resulting shape that does not maintain the desired design intent. We refer to this failure as "reversal constraining".

Morphologic errors in procedural models are summarized in Fig. 5.

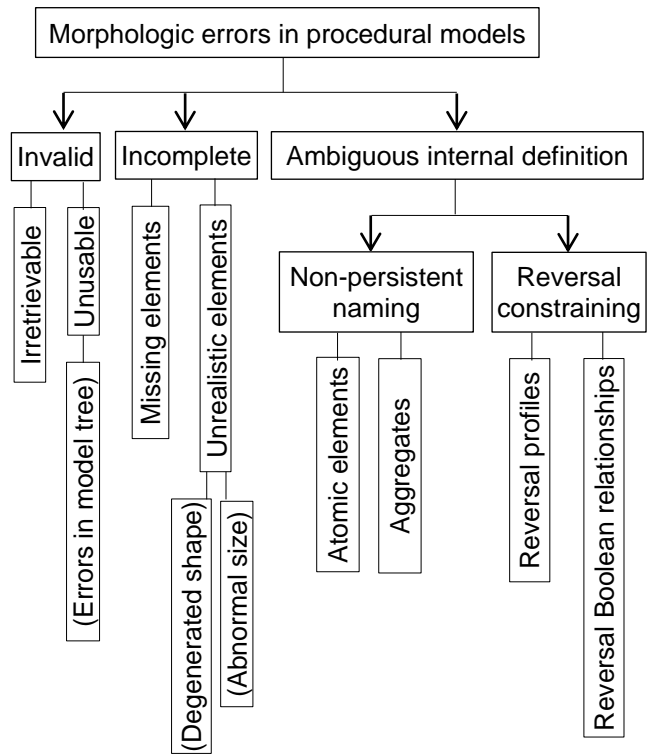

Fig. 5 Morphologic errors in procedural models

### 3.5 Syntactic errors in procedural CAD models

Researchers Tessier and Wang identified two types of data interoperability incompatibilities in procedural CAD systems: *Structural heterogeneity* (caused by the use of different data structures; such as a CAD system that defines a fillet by the removed edge and the radius, while other requires the tangent edges instead of the original edge), and *semantic heterogeneity* (caused by naming and terminology differences, such as a CAD system using the term *fillet* for a feature that other systems recognize as *round*) [78]. While structural heterogeneity is the main cause of morphologic errors (and requires the mapping of features that are equivalent but defined with different data structures), the ultimate goal is the "semantic interoperability" of procedural representations, where the term *semantic* can be broadly defined as the meaning associated with a terminology in a particular context [94].

Direct mapping through geometric modeling kernel-level data exchange (which is a particular type of direct translation that occurs between two different CAD applications that share the same kernel, like ACIS by Dassault Systèmes or Parasolid by Siemens) continues to be an interesting approach for the semantic interoperability of procedural representations [125], but its use is limited to certain types of model exchanges because of its high cost and proprietary nature (the latter is described as a potential drawback in [78]).

Mapping through neutral representations is a more efficient alternative, as procedural models with parameters, features and constraints can be processed.

Although geometric constraint solving has undergone substantial progress in terms of the types of objects and constraints that can be handled robustly, parametric operations have largely remained within the same conceptualization and are now beginning to limit the flexibility of CAD systems [126]. We can classify these limitations through the quality dimensions of consistency and conciseness [115].

A neutral model is *consistent* if it is simultaneously flexible (to enable re-design) and robust (to prevent undesired changes and failures during edition). A neutral model is *concise* if it does not contain repetitive or fragmented constraints, modeling operations or datums, as those used in poor translations when direct mappings are unavailable. A model must also be represented by high level modeling operations (contrary to what happens when they are replaced by low level operations to solve incomplete equivalences between representations).

Mapping cannot maintain attributes that are missing in the original model. For example, *buried features* are those completely included within other features [46]. In the best case scenario, they are an example of overlapping that cannot be mapped as a concise image. In the worst case scenario, they may result in altered models (unexpected geometry and/or topology).

Some authors have used explicit *ontologies* (a formal representation of a set of concepts, their properties, and the relationships between those concepts within a given domain) for the semantic interoperability of CAD systems [78,94,127,128]. Explicit ontologies define representations that are application-independent, expressive and unambiguous [94].

Explicit ontologies are converging with the ISO application protocol for managed model-based 3D engineering, which was recently published as API 242 [129] after an extensive development period [37,93,130].

Dialects or "flavors" are an open problem, as there are no commonly accepted standards on how to solve certain complex geometry that results from combining multiple shapes. A common example involves how a constant-radius blend in an edge should end at complex vertices [78,118].

An interesting new technique to tackle the mapping type problem is the application of the dual model strategy as follows: the primary procedural model is associated to a secondary B-Rep model used by the receiving system to check the validity of the model transfer [48]. This method provides a sort of ground truth, which can be useful. However, there is a large collection of errors that this type of ground truth may contain, as discussed previously. We note that this was not the original aim of dual models, whose origins and current use can be traced back to the work by Kim et al. [131].

Domain problems in procedural models are similar to those described in Section 3.2 for explicit representations. Legacy problems will likely be solved once the new STEP standard becomes fully available and widespread.

Syntactic errors in procedural models are summarized in Fig. 6. Horizontal boxes describe the realm of the errors. Vertical boxes describe common errors linked to those realms.

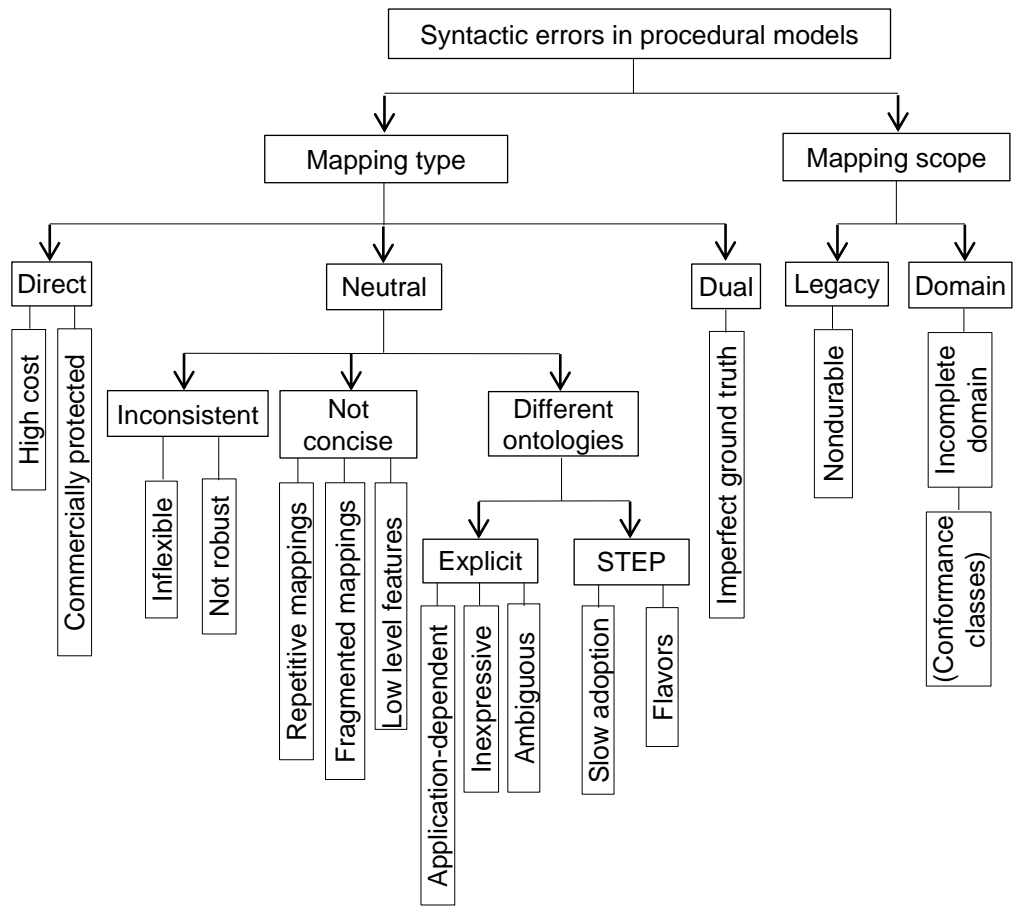

Fig. 6 Syntactic errors in procedural models

### 3.6 Semantic errors in procedural CAD models

Company et al. [115] described two fundamental qualities of a semantic error-free procedural CAD model: the model must be *simple* and *convey design intent*. Poorly perceived models require more time to alter [132]. In this context, the use of formal CAD modeling strategies and best practices for history-based parametric design is growing interest as a method to improve the semantic quality of CAD models. A recent contribution by Camba et al. [133] on approaches specifically designed to emphasize CAD reusability (Delphi's horizontal modeling, explicit reference modeling, and resilient modeling) reveals significant advantages of formal modeling methodologies, particularly resilient techniques, over non-structured approaches, as well as the unexpected problems of the horizontal strategy in numerous modeling situations.

A model is simple if its model tree is *clear* and *understandable*, which means that modeling operations in the modeling tree must be labeled to emphasize their function (instead of how they were built), and related modeling operations must be grouped to emphasize functional parent–child relationships. A model is also simple if it *uses compatible and standard modeling operations.* In fact, standard modeling commands have been proposed as mechanisms to exchange design intent [130].

It has been confirmed that relatively simpler features, the use of reference geometry, and the correct feature sequence are positively correlated with design intent proxy ratings [132]. Nonetheless, the qualities of a model that more directly convey design intent are described as follows [115]:

1. The modeling process must effectively convey the right information about function. Geometric constraints must highlight functional relationships, models must use feature-based operations that convey the functionality of the parts [134], datums must convey the skeleton or scaffold of the model [135], and functional patterns (regularities, symmetries and repetitive patters) must be explicitly included in the model tree [133].
2. The modeling process must be efficacious. It must be free of (1) fragmented operations that mask the final part function, (2) overlapping operations that do not change the model geometry, and (3) overlapping operations that mask previous operations. The model must also be free of generic (unspecific) feature types (i.e., suitable features are not replaced by merely similar ones), and shapes. Also unspecific patterns should be avoided (i.e. those that simply reduce the modeling effort, but do not convey design intent).
3. The model must be efficient. This means that: (1) The model tree distinguishes between datums, scaffolds, core, detail, replication and cosmetic operations; (2) Replication operations (based on patterns) are used to convey functionality in the model tree, and (3) Design decisions are traceable within the model tree.

To a certain extent, some of the previous practices are subjective; sometimes, even contradictory. Therefore, controls are not easily implementable. Nevertheless, some of the recommendations that result in clear, specific, and objective tasks could be easily detected by existing CAD packages. Unfortunately, there is a lack of checking tools to confirm that such rules have been followed. For example, similar to the way most CAD applications highlight insufficiently constrained sketches, they could also highlight default named features in the model tree to encourage users to apply proper naming conventions.

In addition, techniques to simplify secondary models can shed light on the features that should be modeled separately or last, in order to facilitate its removal [16]. However, those techniques have not yet been used to check the quality of the model sequence or the grouping of operations that may be functionally related.

A model is ineffective if its design intent is not maintained when the model is altered. In this regard, knowing the range where topology does not change is important. Authors van der Meiden et al. presented a method to compute the critical values when a single parameter of a model is modified, (i.e. the parameter values for which the topology of the model changes) [136]. Accordingly, MQT tools could simply highlight those parameters that are particularly sensitive to changes, making explicit where a model is robust but barely flexible.

Similar tools such as a consistency checker between the declared typology of the features and its actual nature after editing could also be implemented (e.g., a hole declared as blind at creation that becomes a thru hole after some editing) [137]. Enriching the model with annotations to communicate geometric design intent explicitly is also a related active field of study [138].

Semantic errors in procedural models are summarized in Fig. 7.

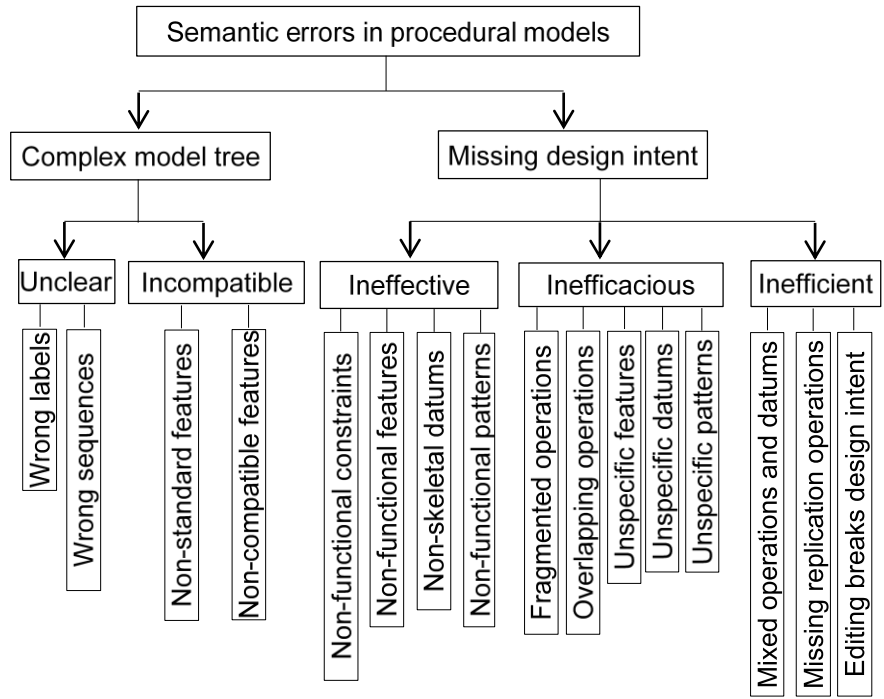

Fig. 7 Semantic errors in procedural models

## 4. Commercial CAD Model Quality Testing Tools

A preliminary analysis of current commercially available MQT's has been summarized in a series of tables, provided in this section. The scopes of the documents generated by CAD applications that can be tested by the various MQT tools available in the market are shown In Table 1. The types of CAD documents that can be typically tested by commercial Quality Testing Tools include models, assemblies, and drawings. Those tools that were unable to test models have been excluded from our study. In our sample, all MQT tools that test models also test assemblies, and most can also test drawings.

Table 1 also shows whether these MQT tools are linked to or embedded into any particular 3D CAD Design software. 36% of the MQT tools studied are embedded.

Table 1. Document checked by different Model Quality Testing tools

| Model Quality Testing tool | | References | Drawings | Models | Assemblies | Embedded system | Linked to a particular CAD |
|---|---|---|---|---|---|---|---|
| | CADfix (ITI) | [49, 139-141] | | Y | Y | | |
| | CADIQ (ITI) | [2, 142-144] | | Y | Y | | |
| | 3DTransVidia | [145] | | Y | Y | | |
| | Cax Quality Manager | [146] | Y | Y | Y | | Siemens NX |
| | Design Checker | [147] | Y | Y | Y | Y | Solidworks |
| | DesignQA® | [148] | | Y | Y | | |
| | GeometryQA® | [149] | Y | Y | Y | | |
| | PrescientQA® | [150] | Y | Y | Y | | CATIA V5 |
| | iCHECK IT | [151] | Y | Y | Y | | CATIA V5, Autodesk Inventor |
| | Knowledge Advisor | [152] | | Y | Y | Y | CATIA |
| | Knowledge Expert | [153] | | Y | Y | Y | CATIA |
| | ModelCHECK | [154] | Y | Y | Y | Y | PTC Creo |
| | NX Check-Mate | [155] | Y | Y | Y | Y | Siemens NX |
| | Q-Checker | [156,157] | Y | Y | Y | | CATIA |

Legend: Y for available document type and void for unavailable.

Average prices of MQT are summarized in Table 2. The table includes three types of MQT tools: embedded systems that obviously require a particular CAD system to be installed; non-embedded systems that are linked to a particular CAD package, and systems that are neither embedded nor linked to any CAD package.

Table 2. List prices of the different Model Quality Testing groups

| Type | Average MQT Price one license (€) | Average Annual maintenance MQT (€) |
|---|---|---|
| Embedded and linked | 1450 | 225 |
| Not-embedded but linked | 6900 | 1150 |
| Neither embedded nor linked | 15500 | 3500 |

Average prices were calculated from list prices that were compiled at the time of writing this paper. They are meant to be indicative, as prices can vary rapidly. Furthermore, list prices may differ from final prices, as sellers often negotiate custom rates with each client. Nevertheless, three price ranges typically stand out: embedded systems which require a particular CAD package to be installed are in the $1,000 to $2,000 range, plus $200-$500 of maintenance; non-embedded systems that are linked to a particular CAD package are in the $5,000-$10,000 range plus $1,000-$3,000 of maintenance, and standalone systems (not linked to any particular CAD package) are in the $10,000-$20,000 range plus $2,000-$5,000 of maintenance.

We conclude that the estimated costs (exact prices are usually unavailable) are high enough to prevent SME's from adopting MQT tools.

Interoperability of MQT tools is illustrated in Table 3. Embedded systems (Design Checker, ModelCHECK, NX Check-Mate and Q-Checker) support less formats than standalone systems, which is logical, as priority is always given to models and assemblies created with their own software. While CATIA files are supported by 71% of the MQT tools, most formats are supported by less than 8% of the MQT tools.

Table 3. Formats supported by Model Quality Testers

| Model Quality Testing tool | AutoCAD | Autodesk | Autodesk Inventor | Autodesk Vault | CATIA | I-deas NX | PTC Creo | Pro/Engineer | Solid Edge | SolidWorks | Unigraphics | ACIS | Parasolid | DXF | DWG | IGES | STEP Unspecific API | STEP AP203 | STEP AP204 | STEP AP242 | CADDS | JT (Open) | NX CAD Application | NX technology | STL | UG-NX | VDAFS | 3D PDF |
|---|---|---|---|---|---|---|---|---|---|---|---|---|---|---|---|---|---|---|---|---|---|---|---|---|---|---|---|---|
| CADfix (ITI) | | | Y | | Y(3) | | | Y | | Y | | Y | Y | Y | Y | Y | | Y | Y | Y | Y | Y | | | Y | Y | Y | |
| CADIQ (ITI) | | | Y | | Y(2) | Y | Y | Y | Y | | Y | Y | Y | Y | Y | Y | | | | | | Y | | Y | | | | Y |
| 3DTransVidia | | Y | | | Y(3) | Y | Y | | Y | Y | Y | Y | | | Y | Y | | | | | Y | Y | | | Y | Y | | |
| Cax Quality Manager | | | | | | | | | | | | | | | | | | | | | | | Y | | | | | |
| Design Checker | | | | | | | | | | Y | | | | | Y | | | | | | | | | | | | | |
| DesignQA® | | | | | Y(1) | Y | | Y | | | Y | | | | | | | | | | | | | | | | | |
| GeometryQA® | | | | | Y(1) | Y | | Y | | | Y | | | | | | | | | | | | | | | | | |
| PrescientQA® | | | | | Y(2) | Y | | Y | | | Y | | | | | | | | | | | | | | | | | |
| iCHECK IT | Y | Y | Y | Y | Y(2) | | | | | | | | | | | | | | | | | | | | | | | |
| Knowledge Advisor | | | | | Y | | | | | | | | | | | | | | | | | | | | | | | |
| Knowledge Expert | | | | | Y | | | | | | | | | | | | | | | | | | | | | | | |
| ModelCHECK | | | | | | | Y | | | | | | | | | | | | | | | | | | | | | |
| NX Check-Mate | | | | | | | | | | | | | | | | | | | | | | | | Y | | | | |
| Q-Checker | | | | | Y | | | | | | | | | | | | | | | | | | | | | | | |

Legends: Y for available format and "blank" for unavailable. [1]CATIA V4 system interface, [2]CATIA V5 system interface, [3]CATIA V4/V5 system interface

We have reviewed the model quality tools listed in Tables 1 and 3 in light of the criteria defined in the taxonomy (this is, those defined in Figures 2 to 7). The procedure consisted in mapping the taxonomy criteria against the requirements that the tools can check. As an example, the mapping for the module Build Checks of SolidWorks Design Checker— a tool to set the requirements for evaluation—is shown in Table 4. The requirements that this module is able to check are grouped in up to 7 different categories: document, annotations, dimension, drawing document, part document, assembly document and feature. Obviously, requirements for groups *drawing document* and *assembly document* are not related to model quality, so these requirements were excluded from the mapping. Alternatively, *part document* and *feature* checks are clearly related. In addition, some *document*, *annotations* and *dimension* checks were also included in the mapping, as they are more or less transversal to models, drawings and assemblies. Certain requirements of those groups, however, were excluded as they are limited to drawings (Table font, Balloon font, References up to date, etc.). Finally, some requirements were only indirectly linked to the taxonomy. Material property is used for drawings' hatch sections and to calculate mass properties. Therefore, an incorrect material selection may result in a non-standard hatching, which, in turn, results in an inconsistent standardization. Furthermore, the selection can also produce an incorrect calculation of mass properties, which may misguide a geometrical accuracy problem. A missing blend table results in undetectable blend constraints that, in turn, prevent its re-featuring.

Table 4. Model quality criteria supported by the module Build Checks of SolidWorks Design Checker

| Group | Requirement | Type of error (and figure where it appears in the taxonomy) | Explicit | | | Procedural | | |
|---|---|---|---|---|---|---|---|---|
| | | | Morphologic | Syntactic | Semantic | Morphologic | Syntactic | Semantic |
| Part Document Checks | Material check | Geometrical accuracy problem, or sectorial standardization (Fig. 3) | | * | | | | |
| | Blend table | Undetectable constraints (Fig. 4) | | | * | | | |
| Feature Checks | Feature error warnings | Incomplete (Fig. 5) | | | | * | | |
| | Fully defined sketch | Inconsistent, not robust (Fig. 6) | | | | | * | |
| | Standard hole size | Undetectable features (Fig. 4) | | | * | | | |
| | Suppressed feature | Complex model tree, or missing design intent (Fig. 7) | | | | | | * |
| | Feature positioning check | Ambiguous internal definition (Fig. 5) | | | | * | | |
| Document Checks | Dimensioning standard | Sectorial standardization (Fig. 3) | | * | | | | |
| | Arrow style document check | Sectorial standardization (Fig. 3) | | * | | | | |
| | Custom property | Sectorial standardization (Fig. 3) | | * | | | | |
| | Units setting | Sectorial standardization (Fig. 3) | | * | | | | |
| | Surface finish font | Sectorial standardization (Fig. 3) | | * | | | | |
| | Weld symbol Font | Sectorial standardization (Fig. 3) | | * | | | | |
| | Dimension Font | Sectorial standardization (Fig. 3) | | * | | | | |
| | Overridden mass check | Geometrical accuracy problem, or sectorial standardization (Fig. 3) | | * | | | | |

| | Virtual Sharp | Undetectable constraints (Fig.4) | | | * | | | |
|---|---|---|---|---|---|---|---|---|
| Annotation Checks | Arrow style | Sectorial standardization (Fig. 3) | | * | | | | |
| | Font style | Sectorial standardization (Fig. 3) | | * | | | | |
| | GTOL Datum | Missing design intent, unspecific datums (Fig. 7) | | | | | | * |
| | Spell Checker | Sectorial standardization (Fig. 3) | | * | | | | |
| Dimension Checks | Overridden dimension | Sectorial standardization (Fig. 3) | | * | | | | |
| | Arrow style dimension check | Sectorial standardization (Fig. 3) | | * | | | | |
| | Font style | Sectorial standardization (Fig. 3) | | * | | | | |
| | Units setting | Sectorial standardization (Fig. 3) | | * | | | | |
| | Overlapping dimension | Undetectable constraints (Fig.4) | | | * | | | |
| | Replaced original text | Sectorial standardization (Fig. 3) | | * | | | | |
| | Text position | Sectorial standardization (Fig. 3) | | * | | | | |
| | Dimension precision | Sectorial standardization (Fig. 3) | | * | | | | |

As a result of the mappings described earlier, a summary of the performance levels of each MQT tool is provided in Table 5 for each of the six topics of the taxonomy defined in Section 3. Preliminary conclusions include:

- The morphological level is reasonably well covered in explicit representations, but there is still room for improvement.
- The syntactic quality in explicit representations can be better improved by using efficient modeling strategies, rather than by healing poorly mapped translations.
- Procedural representations have not yet been thoroughly considered by MQT's.

Table 5. Semantic levels supported by the Model Quality Testers

| | | Explicit | | | Procedural | | |
|---|---|---|---|---|---|---|---|
| | | Morphologic | Syntactic | Semantic | Morphologic | Syntactic | Semantic |
| Model Quality Tool | CADfix (ITI) | *** | *** | *** | | | |
| | CADIQ (ITI) | *** | ** | *** | ** | ** | ** |
| | 3D TransVidia | ** | *** | *** | *** | ** | *** |
| | Cax Quality Manager | * | | * | | | |
| | Design Checker | | ** | * | | | |
| | DesignQA® | * | * | * | | | |
| | GeometryQA® | ** | * | ** | | | |
| | PrescientQA® | ** | * | ** | | | |
| | iCHECK IT | ** | * | *** | | | |
| | KnowledgeAdvisor | | | * | | | |
| | KnowledgeExpert | * | | * | | | |

| | | | | | | |
|---|---|---|---|---|---|---|
| ModelCHECK | *** | | ** | *** | | ** |
| NX Check-Mate | ** | | *** | | | |
| Q-Checker | *** | | *** | *** | | *** |

Legend: *** high coverage, ** average coverage, * poor coverage, and "blank" for no coverage.

The tables included in this section summarize our qualitative evaluation of the information that is currently available for commercial MQT tools. The limited availability of information and the lack of homogeneity in terms of the advertised details of each tool may have an effect on the results. In addition, identical parameters or tools are called differently depending on the system. To the best of our knowledge, a quantitative evaluation that compares each MQT tool against a common benchmark has not yet been conducted. The metrics used in our approximation are defined as follows:

- The available information on geometric or topological detections for each MQT was used to determine the coverage of morphological failures in explicit and procedural representations.

- The main factor to determine syntactic failures in explicit and procedural representations was the number of supported formats, as indicated in the product's documentation. IGES and STEP were included as complementary formats.

- Two criteria were considered to evaluate coverage of semantic parameters in explicit and procedural representations. First, whether the MQT repaired a detected error (and whether this procedure was manual or automatic); and second, the amount of errors detected was used to determine a qualitative score of high, average, or low coverage.

Our tables are only intended to provide a broad picture of the current state-of-the-art in MQT systems. Although conclusions must be validated through quantitative studies, we can report that even though a number of commercial MQT tools are available, the technology is not widespread. Reasons include: the proprietary nature of most tools (they only work as part of a particular CAD application), their high cost, and the fact that they are not comprehensive (only a fraction of the inconsistencies, inaccuracies and failures of CAD models can be detected).

## 5. Open problems

By analyzing current MQT tools in the context of our taxonomy, we identified some aspects that require further study and development.

Different types of quality loss are associated with master model changes, and also with its representation type. In short, simplification may produce non representative secondary models if design intent—which should guide the simplification—is absent in the master model. Mapping procedures aimed at guaranteeing interoperability will always be prone to data corruption. Finally, CAD model reuse is particularly sensitive to hidden errors and anomalies. Design intent is still poorly addressed by MQT tools, but there are other bottlenecks for different stakeholders (Original Equipment Manufacturers (OEM), lower tier suppliers, and SME's).

Dominant OEM's force top-down interoperability with their suppliers, which results in "defensive" or "conservative" designs, which are robust but hardly creative. Interoperability is a main concern for OEM's, as reusability is guaranteed by the best practices they impose, whereas simplification tasks are transferred to suppliers. A hidden problem that hinders interoperability is the lack of proven modeling guidelines [125]. Best practices are checked by MQT tools but also imposed and tuned by the OEM, whose current goal involves improving interoperability by abandoning explicit representations and adopting STEP AP 242. Validation rules require setup, and although quantitative metrics for shape errors already exist, they are context dependent and are governed by computational threshold values that are different for each MQT tool [8]. Better quantitative metrics are still required.

Lower tier suppliers are mainly involved with CAM/CAE tasks. Therefore, they need to improve simplification to reduce costs. Therefore, they must introduce strategic modeling by distinguishing main and secondary features. They also need to remove intermediate formats (for example, by producing NC directly from the master model), or even eliminate secondary views entirely [158], which would increase the need for a good quality master model even further.

For SME's, the demand for interoperability is low. Usually, there is direct contact between the original designer and the user in need of model reuse, and suppliers do not send back their secondary views to the main SME. Alternatively, SME's need to improve reusability. They must be taught how to strategically use CAD, as they do not learn it through best practices (best practices are not imposed by any OEM). SME's also require low cost MQT tools (both in terms of acquisition and maintenance), which should not be tuned by or linked to alien best practices. While automotive and aerospace companies are successfully implementing CAD-based Virtual Prototyping for optimizing processes, small and medium-sized enterprises (SME) frequently fail because of the lack of necessary preconditions [4]. Commercially available MQT tools are expensive, and thus rarely adopted by SME's.

## 6. Conclusions

Quality testing tools for model verification, validation, and comparison are essential, as exporting CAD models that contain errors or anomalies to different downstream applications is prone to data corruption, which typically requires the models to be reworked by the downstream user.

In this paper, a new taxonomy of issues related to CAD model quality was defined. It was validated by using it to classify the currently available CAD quality assurance tools and determine which aspects of quality are reasonably addressed, and which remain open problems.

The new taxonomy is based on the assumption that classifying only by representation type or only by the semantic level of the failures are both incomplete views. The proposed taxonomy distinguishes between explicit and procedural models, and, for each type of model, morphologic, syntactic, and semantic errors are characterized. Additionally, each semantic level has been paired to a particular type of change: morphologic to simplification, syntactic to interoperability, and semantic to reuse. This pairing is not strict, as some crossed relations still occur.

We hypothesize that taxonomies of such a complex and evolving subject will always be incomplete, but useful. By using our resulting frame, we have shed light on the different states-of-the-art in explicit and procedural models. For explicit models, we have argued the following:

(1) Morphological correctness of explicit models can be evaluated with ad hoc software.

(2) Interoperability is minimally supported by suitable standards such as STEP AP 203.

(3) Efficient and adaptive model reuse is unfeasible for purely explicit representations, but limited for enriched explicit representations.

With regards to procedural models, we have argued that:

(1) Interactive modeling editors prevent most morphologic errors, and MQT tools can handle the remaining reasonably well.
(2) Interoperability problems between procedural representations are likely to improve drastically with the development of STEP AP242 (although implementation and adoption have been slow).
(3) Higher semantic aspects of quality—such as assurance of the design intent embedded in the master model—are hardly addressed by current CAD quality testers.

Although the scope and level of detail of the comparison of current MQT tools limits our study, two preliminary conclusions arise:

(1) MQT tools are mostly aimed at homogenizing the vast amount of documents produced and shared by large OEM's, and thus are primarily aimed at preventing easily solvable low-semantic level mistakes and incoherencies.
(2) MQT tools are still unaffordable for many Small and Medium Enterprises, as they are expensive both in terms of cost and training time. We presume that they will only become valuable for this particular market segment if document homogenization ceases to be prevalent over conveying design intent.

Although valid as plausible hypotheses, it can be argued that an in-depth analysis of all commercially available MQT tools would be necessary to fully support our conclusions. Nonetheless, the most significant contribution of this work is the value and effectiveness of the taxonomy, as demonstrated by the classification of strengths and weaknesses of MQT tools.

As a future development, a more comprehensive theoretical basis is required to define quantitative metrics for complex quality requirements. Although some metrics for shape errors already exist (such as those considered in Part 42 of STEP), further improvement is required, as these metrics are context dependent and require tuning by expert users.

Finally, higher level semantics in the master model are required to guarantee that *design intent* is made explicit and easily understandable by all users involved in creating downstream models (for example, was a symmetric part inadvertently made non-symmetric during the simplification process required to create a FE mesh?).

**Acknowledgements**

This work was supported by the Spanish Ministry of Economy and Competitiveness and the European Regional Development Fund, through the ANNOTA project (Ref. TIN2013-46036-C3-1-R). The authors also wish to thank the editor and reviewers for their valuable comments and suggestions that helped us improve the quality of the paper.